\documentclass{ws-procs9x6}

\newcommand{\bra}[1]{\left\langle #1 \right|} 
\newcommand{\ket}[1]{\left| #1 \right\rangle}

\newcommand{\ubar}{\overline{u}}
\newcommand{\dbar}{\overline{d}}

\begin{document}

\title{Light-Cone Sum Rules for the Form Factors of the $N\gamma\Delta$ transition at $Q^2=0$}

\author{J. Rohrwild$^*$}

\address{Institut f\"ur Theoretische Physik, Regensburg University,\\
Regensburg, D-93040, Germany\\
$^*$E-mail: juergen.rohrwild@physik.uni-r.de\\
}

\begin{abstract}
The radiative $\Delta \to \gamma N$ transition is examined at the real photon point $Q^2=0$ using the framework of light-cone QCD sum rules. In particular, we determine the sum rules for the transition form factors $G_M(0)$ and $R_{EM}$ up to twist 4.
\end{abstract}

\keywords{hard exclusive processes, distribution amplitudes, form factors}

\bodymatter

\section{Introduction}

\begin{figure}
\begin{center}
\epsfig{figure=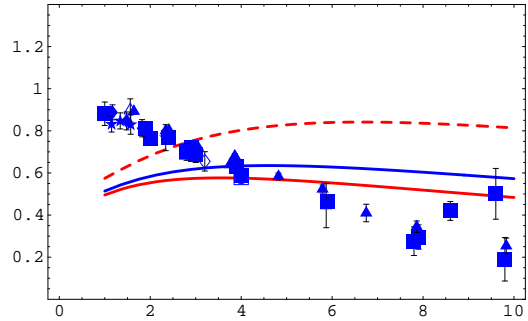, width=0.6 \textwidth}
\end{center}
\caption{LCSRs for the $\gamma^{*} N \to \Delta$ transition form factor $G_{M}(Q^2)/(3G_{\rm Dipole})$ vs. $Q^2$  from [17]. The different lines correspond to different shapes of the nucleon distribution amplitude. For references to the experimental data see [17].\label{BLPR}}
\end{figure}

In this talk we present a light-cone sum rule (LCSR) calculation for the $N\gamma \Delta$ transition form factors at $Q^2=0$. The form factors are by themselves interesting quantities as they contain information on the baryon structure. Especially, it is known that a non-vanishing electromagnetic quadrupole transition form factor $G_E(0)$ indicates a deformation of the nucleon-$\Delta$ system \cite{Glashow:1979gp}. Therefore, the $N\gamma \Delta$ transition has been studied extensively by both experimentalists \cite{Frolov:1998pw,Joo:2001tw,Kunz:2003we,Ungaro:2006df,Sparveris:2006uk,Blanpied:2001ae,Beck:1999ge,Sandorfi:1998xr,Beck:1997ew} and theorists \cite{Isgur:1981yz,Wirzba:1986sc,Pascalutsa:2005vq,Alexandrou:2004xn,Pascalutsa:2006up}.\\
Our calculation is primarily motivated by the fact that an early SVZ sum rule-based examination of the transition did not yield a definite result \cite{Ioffe:1983ju}. Moerover, a recent light-cone sum rule calculation\cite{Braun:2005be} based on nucleon distribution amplitudes (DAs) \cite{Braun:2000kw} shows a notable deviation of the magnetic dipole form factor $G_M(Q^2)$ for $1\;{\rm GeV}^2 <Q^2<2\;{\rm GeV}^2$ from experiment, see Fig.\ref{BLPR}, while this approach works perfectly for the nucleon electromagnetic form factors \cite{Braun:2006hz,Braun:2001tj,Lenz:2003tq}. The reason for this discrepancy is not yet understood and the presented analysis is a first step towards a better understanding. Different LCSR analyses of the $N\gamma \Delta$ transition at $Q^2=0$ can be found in \cite{Aliev:1999tq} and, as part of a more general analysis, in \cite{Aliev:2004ju}.\\
In the following, we will give a short introduction to LCSR calculations with photon DAs using the example of the proton magnetic moment and  present the results for the form factors of the $N\gamma \Delta$ transition at $Q^2=0$.

\section{Light-cone sum rules for form factors}

The starting point for the LCSR approach at $Q^2=0$ is a correlation function of the form
\begin{equation}
\Pi \left(p,q \right) = i^2  \int\! d^4 x  \int \!d^4 y \;e^{i px + i qy} \bra{0} \mathcal{T} \lbrace \eta' \! \left( x\right) j^{\mu}_{\rm em}\left( y \right)\overline{\eta}\! \left( 0 \right) \rbrace  \ket{0} e_{\mu}^{(\lambda)}\mbox{ ,}
\end{equation}
which describes the transition of a baryon $B(p+q)$ to a baryon $B'(p)$ by an interaction with the electromagnetic current $j_{\rm em}^{\mu}(y)$ ($e_{\mu}^{(\lambda)}$ is the 4-polarization vector of the photon). The baryons are created by three-quark currents $\eta'$ and $\eta$, respectively.

For simplicity, we will consider the explicit example of the magnetic moment of protons, which has been considered e.g. in \cite{Braun:1988qv}. The interpolating fields $\eta$ and $\eta'$ are replaced by the Ioffe-current for the proton
\begin{equation}
\label{corr}
 \eta_{\rm Ioffe}(x)= \left(u^a(x)\mathcal{C}\gamma^{\lambda} u^{b}(x) \right) \gamma_5 \gamma_{\lambda }d^{c}(x)\varepsilon^{abc}
\end{equation}
and the electromagnetic current can be written in the from 
\begin{equation}
j^{\mu}_{\rm em}(y)=e_d \dbar(y) \gamma_{\mu} d(y) + e_u \ubar(y) \gamma_{\mu} u(y).
\end{equation}
The standard strategy of the light-cone sum rule approach involves the calculation of the correlation function (\ref{corr}) in two different regimes:
\begin{itemize}
\item on the hadron level the correlator is expressed in terms of form factor (in case of the proton the Dirac and Pauli form factors $F_1$ and $F_2$)
\item on the level of quarks the correlation function is expanded in terms of photon distribution amplitudes of definite twist.
\end{itemize}
As the interpolating field $\eta$ has non-vanishing overlap with states of higher mass, whose quantum numbers coincide with those of the current $\eta$, Eq.(\ref{corr}) contains  contribution of all these states. After equating both hadronic and quark representation a Borel transformation can be employed to suppress the higher mass states exponentially. As the momenta $p$ and $p+q$ are independent, it is possible to perform a so-called double Borel transformation and introduce two Borel parameters $M_1$ and $M_2$, corresponding to $p$ and $p+q$. Introducing a model for the excited states and the continuum one can subtract these contributions from both sides and express the form factors via the photon distribution amplitudes\footnote{The distribution amplitudes are non-perturbative in nature and can be accessed e.g. via QCD sum rule estimates\cite{Balitsky:1989ry,Ball:2002ps}, LCSRs \cite{Rohrwild:2007yt} or lattice calculations.}, which are known up to twist 4 \cite{Ball:2002ps}. Besides the dependence on $M_1$ and $M_2$ the form factors are also functions of the continuum subtraction threshold $S_0$.\\

\begin{figure}
\begin{center} 
\epsfig{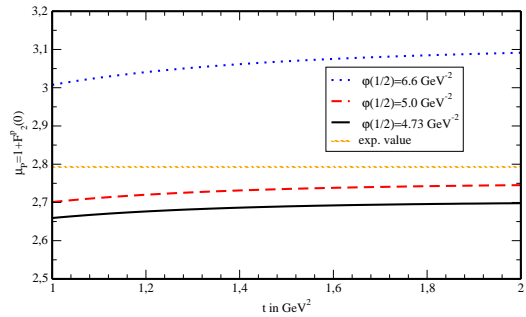}
\end{center}
\caption{Magnetic moment of the proton $\mu_p$ vs. Borelparameter $t=(M_1^2+M_2^2)/M_1^2M_2^2$ for different central values of the photon wave function $\varphi(1/2)$ (Plot from [29]). The solid line is obtained by the choice of an asymptotic shape and a magnetic susceptibility of the quark condensate of $\chi(\mu=1 {\rm GeV}^2)=3.15 \pm 0.3 {\rm GeV}^{-2}$ [26]. The hatched (orange) line represents the experimental value .\label{mup}}
\end{figure}
The process $p\to p \gamma$ is inherently symmetric in the initial and final hadron state. Therefore, it is natural to choose both Borel parameters equal: $M_1^2=M_2^2$. With this choice the leading-twist photon wave function at the middle point $\varphi(1/2)$, see e.g. \cite{Braun:1988qv,Belyaev:1994zk}, is  the main non-perturbative input parameter. The result for the magnetic moment of the proton for different values of $\varphi(1/2)$ is shown in Fig.\ref{mup}. The agreement with experiment is very good.

\section{Nucleon-$\Delta$-transition}

The radiative $\Delta\to N$ decay can also be described by the correlation function \eqref{corr}, the current $\eta'$ has to be replaced by 
\begin{equation}
\label{deltacurrent}
\eta^{\mu}_{\Delta}(x)= \left[\left(u^a(x)\mathcal{C}\gamma^{\mu} u^{b}(x) \right) d^{c}(x)+2\left(u^a(x)\mathcal{C}\gamma^{\mu} d^{b}(x) \right) u^{c}(x) \right]\varepsilon^{abc} \mbox{ .}
\end{equation}
Compared to the standard approach presented in the previous section, several subtleties have to be taken into account.
First and foremost, the interpolating field \eqref{deltacurrent} is known to have non-vanishing overlap with states of spin $1/2$ and negative parity, these have to be removed ``by hand''  by a specific choice of the Lorentz basis before the continuum subtraction is performed, see e.g. \cite{Rohrwild:2007iz,Belyaev:1995uw} for two possible choices. Furthermore, it has been argued that a symmetric choice of the two Borel parameters, $M_1^2=M_2^2$, is not optimal. In order to take into account the mass difference of $300\; {\rm MeV}$ between the nucleon and the $\Delta$ the choice of $M_1^2/M_2^2=m_{\Delta}^2/m_{N}^2$ is advantageous. This  strategy was advocated in \cite{Balitsky:1989ry}. \\
\begin{figure}
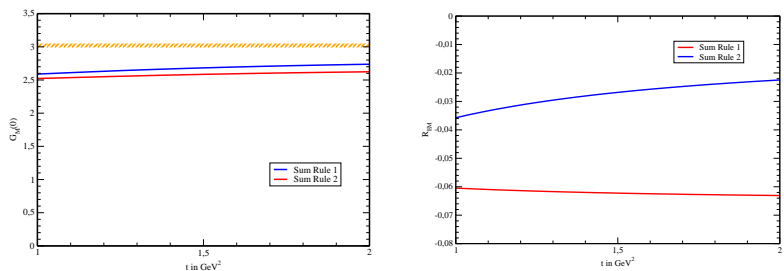

\begin{minipage}{0.47\textwidth}
\epsfig{figure=GM2plotforSiegen.eps, width=0.9 \textwidth}
\end{minipage}
\begin{minipage}{0.47\textwidth}
\epsfig{figure=REMplotforSiegen.eps, width=0.9 \textwidth}
\end{minipage}
\caption{Light-cone sum rules for the $N \gamma \Delta$ transition form factors (left panel: $G_M(0)$ vs. $t$ ; right panel: $R_{EM}$ vs. $t$) in each case obtained from two different Lorentz structures. The experimental value is given by the hatched region.\label{GM}}
\end{figure}

The result for the magnetic dipole form factor $G_M(0)$ is shown in the left panel of Fig.\ref{GM}. The two solid lines correspond to two different sum rules that stem from different Lorentz structures. Within errors, the sum rule prediction \cite{Rohrwild:2007iz}
\begin{equation}
G_{M}(0)=2.70 \pm 0.27
\end{equation}
agrees well with the experimental value of $G_{M}(0)=3.03 \pm 0.03$ \cite{Frolov:1998pw}, which is displayed by the hatched region. 
The right panel in Fig.\ref{GM} shows the ratio of the electric quadrupole form factor $G_{E}(0)$ and the  magnetic dipole form factor $G_M(0)$, $R_{EM}:=G_E(0)/G_M(0)$. The solid lines again correspond to sum rules for different Lorentz structures. Although the upper line seems to agree very well with the value of $R_{EM}=-2.5\pm0.4 \%$ given by PDG \cite{Yao:2006px}, this is actually misleading. Not every structure is suitable for a sum rule and, indeed, the lower line, which yields 
\begin{equation}
  R_{EM}(0)=-(6.4 \pm 0.8)\%
\end{equation}
corresponds to the more ``trustworthy'' sum rule \cite{Rohrwild:2007iz}. This is further supported by its marginal dependence on the unphysical Borel parameter $t=(M_1^2+M_2^2)/M_1^2 M_2^2$. Our result agrees with experiment only within a factor $2$, however, it should be pointed out that the smallness of $R_{EM}$ is largely due to cancellations. The quantity is therefore intrinsically difficult to access via a sum rule.

\section{Conclusions}

Photon distribution amplitudes provide an intriguing device to treat radiative decays within the method of LCSRs and allow the determination of transition from factors at $Q^2=0$, which are usually not accessible when using baryon distribution amplitudes. Our result for $\mu_p$ is in very good agreement with experiment and lends further support to this method. \\
$G_M(0)$ is rather close to the corresponding experimental value. While this is desirable, it does not shed much light on the peculiar behavior of LCSR predictions for large values of $Q^2$ \cite{Braun:2005be}, which are almost a factor $2$ below data for momentum transfers below $2 \; {\rm GeV}^2$. However, our result indicates that the effect observed in \cite{Belyaev:1995uw,Braun:2005be} is probably not due to the choice of the interpolating field. In order to close the gap to aforementioned calculation, it is necessary to expand our approach from the real photon point to virtualities ranging from $0$ to $-1 {\rm GeV}^2$. This requires photon distribution amplitudes for virtual photons, see e.g. \cite{Yu:2006fq}.\\
Moreover, it would be desirable to have more precise values for the non-perturbative input parameters appearing in the photon distribution amplitudes, since e.g. the parameters of twist 4 are only known up to $50\%$. 

\section*{Acknowledgements}

I would like to thank V. Braun and A. Lenz for many enlightening discussions and the organizers of this workshop for their perfect work.

\bibliographystyle{ws-procs9x6}
\bibliography{NewportProceedingRohrwild}

\end{document}